\documentclass[english]{article}
\usepackage[T1]{fontenc}
\usepackage[latin9]{inputenc}
\usepackage{float}
\usepackage{amssymb}
\usepackage{stackrel}
\usepackage{graphicx}
\usepackage{esint}

\makeatletter

\pdfoutput=1
\usepackage{amsfonts,amssymb}
\usepackage{latexsym}
\usepackage{epsfig}
\usepackage{cite}
\usepackage{graphicx}
\usepackage[colorlinks,linkcolor=blue]{hyperref}
\newcommand{\be}{\begin{equation}}
\newcommand{\ee}{\end{equation}}

\makeatother

\usepackage{babel}
\begin{document}
{}~ \hfill\vbox{\hbox{hep-th/yymmnnn}}\break
\vskip 3.0cm
\centerline{\Large \bf Extremal black string with Kalb-Ramond field via $\alpha^{\prime}$ corrections}

\vspace*{10.0ex}
\centerline{\large Shuxuan Ying}
\vspace*{7.0ex}
\vspace*{4.0ex}
\centerline{\large \it Department of Physics, Chongqing University}
\vspace*{1.0ex}
\centerline{\large \it Chongqing, 401331, China} \vspace*{1.0ex}
\vspace*{4.0ex}

\centerline{ysxuan@cqu.edu.cn}
\vspace*{10.0ex}
\centerline{\bf Abstract} \bigskip \smallskip
In this paper, we obtain the three-dimensional regular extremal black string solution incorporating $\alpha'$ corrections and a non-trivial Kalb-Ramond field. The difficulty in considering the Kalb-Ramond field lies in the fact that it transforms the original equations of motion into an infinite summation form involving matrices, making it difficult to calculate the matrix differential equations. To solve this problem, we introduce a new method that transforms the infinite summation of  matrix differential equations into a simple trace of the matrix. As a result, we are able to obtain a non-perturbative and non-singular extremal black string solution. Indeed, this work serves as a good example for studying more complicated non-perturbative solutions that incorporate the Kalb-Ramond field via complete $\alpha'$ corrections.

\vfill \eject
\baselineskip=16pt
\vspace*{10.0ex}
\tableofcontents

\section{Introduction}

In recent studies \cite{Hohm:2015doa,Hohm:2019ccp,Hohm:2019jgu},
Hohm and Zwiebach made notable contributions to understanding the
$\alpha^{\prime}$ corrections of closed string theory at all orders
and derived the complete action by utilizing the $O\left(d,d\right)$
symmetry. The key point that complete action can be obtained by $O\left(d,d\right)$
symmetry is due to the fact that when massless closed string fields
are independent of the $m$ coordinates, the action manifests an $O\left(m,m\right)$
symmetry \cite{Sen:1991zi,Sen:1991cn}. This result has been verified
in the tree-level string effective action as well as its first-order
$\alpha^{\prime}$ correction. Consequently, it suggests that the
low energy effective action with complete $\alpha^{\prime}$ corrections
can be expressed by utilizing the standard $O\left(d,d\right)$ matrix
in terms of $\alpha^{\prime}$ corrected fields for time-dependent
backgrounds \cite{Veneziano:1991ek, Sen:1991zi,Sen:1991cn, Meissner:1991zj,Meissner:1996sa}.
Based on this observation, Hohm and Zwiebach conjectured that this
result is always true for all orders in $\alpha^{\prime}$. Conjectured
this assumption, they successfully demonstrated that the $\alpha^{\prime}$
corrections can be classified based on even powers of the Hubble parameter
within the framework of the FLRW cosmological background. In addition,
it is worth noting that the dilaton field only includes first-order
time derivatives. These features make the equations of motion (EOM)
to solely include two derivatives of the metric and be exactly solvable.
This remarkable progress introduces a new action that make the study
of black holes and cosmology incorporating non-perturbative string
effects possible. By using this action, numerous new regular black
hole solutions and cosmological solutions were studied. Refs. \cite{Wang:2019kez,Wang:2019dcj,Wang:2019mwi,Wang:2020eln,Gasperini:2023tus,Song:2023txa}
have considered the FLRW cosmological ansatz, revealing that the $\alpha^{\prime}$
corrections can eliminate the big bang singularity. In addition, the
resolution of curvature singularities in two-dimensional string black
holes has been discussed in refs. \cite{Ying:2022xaj,Ying:2022cix,Codina:2023fhy,Ying:2023gmd}.
However, these solutions impose significant constraints on the form
of the spacetime ansatz and require the Kalb-Ramond field to vanish.
It looks impossible to get the non-perturbative solutions with a non-constant
Kalb-Ramond field. The challenge arises from the fact that considering
the Kalb-Ramond field makes the classification of $\alpha^{\prime}$
corrections complicated, which can no longer be simply written by
even powers of the Hubble parameter. Instead, an infinite summation
of $O\left(d,d\right)$ matrices must be considered. The previous
discussion about Kalb-Ramond field in Hohm-Zwiebach action can be
found in a ref. \cite{Bernardo:2021xtr}. To demonstrate the difficulty
including Kalb-Ramond field, let us recall the Hohm-Zwiebach action
for the black hole background:

\begin{equation}
I_{HZ}=-\int dx^{3}e^{-\Phi}\left(-\dot{\Phi}^{2}+\stackrel[k=1]{\infty}{\sum}\left(\alpha^{\prime}\right)^{k-1}\bar{c}_{k}\mathrm{Tr}\left(\dot{\mathcal{S}}^{2k}\right)\right),\label{eq:1 action}
\end{equation}

\noindent where all fields depend on $y$, and $\dot{f}\left(y\right)\equiv\partial_{y}f\left(y\right)$.
The coefficients are $\bar{c}_{1}=c_{1}=-\frac{1}{8}$, $\bar{c}_{2}=-c_{2}=-\frac{1}{64}$,
$\bar{c}_{3}=c_{3}=-\frac{1}{3.2^{7}}$, $\bar{c}_{4}=-c_{4}=-\frac{1}{2^{15}}+\frac{1}{2^{12}}\zeta\left(3\right)$
and $\bar{c}_{k>4}$'s are unknown coefficients for the bosonic case
\cite{Codina:2021cxh}. The $O\left(d,d\right)$ invariant dilaton
and $O\left(d,d\right)$ matrix are defined as

\begin{equation}
e^{-\Phi}\equiv\sqrt{g}e^{-2\phi},\qquad\mathcal{S}=\left(\begin{array}{cc}
BG^{-1} & G-BG^{-1}B\\
G^{-1} & -G^{-1}B
\end{array}\right),
\end{equation}

\noindent where $\phi$ is the physical dilaton, and $g$ does not
include the direction of $dy$. The EOM derived from the action are
given by:

\begin{eqnarray}
\ddot{\Phi}-\stackrel[k=1]{\infty}{\sum}\left(\alpha^{\prime}\right)^{k-1}k\bar{c}_{k}\mathrm{Tr}\left(\dot{\mathcal{S}}\right)^{2k} & = & 0,\label{eq:EOM1}\\
\frac{d}{dy}\left(e^{-\Phi}\stackrel[k=1]{\infty}{\sum}\left(\alpha^{\prime}\right)^{k-1}4k\bar{c}_{k}\mathcal{S}\left(\dot{\mathcal{S}}\right)^{2k-1}\right) & = & 0,\label{eq:EOM2}\\
\dot{\Phi}^{2}-\stackrel[k=1]{\infty}{\sum}\left(\alpha^{\prime}\right)^{k-1}\left(2k-1\right)\bar{c}_{k}\mathrm{Tr}\left(\dot{\mathcal{S}}\right)^{2k} & = & 0.\label{eq:EOM3}
\end{eqnarray}

\noindent If we consider the three-dimensional black hole ansatz without
the Kalb-Ramond field:

\begin{equation}
ds^{2}=a\left(y\right)^{2}\left(-dt^{2}+dx^{2}\right)+dy^{2}.
\end{equation}

\noindent The EOM (\ref{eq:EOM1}) to (\ref{eq:EOM3}) can be significantly
simplified as follows:

\begin{eqnarray}
\dot{\Phi}+\frac{1}{2}\bar{H}\bar{f}\left(\bar{H}\right) & = & 0,\nonumber \\
\frac{d}{dy}\left(e^{-\Phi}\bar{f}\left(\bar{H}\right)\right) & = & 0,\nonumber \\
\dot{\Phi}+\bar{g}\left(\bar{H}\right) & = & 0,\label{eq:EOM}
\end{eqnarray}

\noindent where the Hubble parameter is defined as $\bar{H}\left(y\right)=\frac{\dot{a}\left(y\right)}{a\left(y\right)}$
and

\begin{eqnarray}
\bar{f}\left(\bar{H}\right) & = & d\stackrel[k=1]{\infty}{\sum}\left(-\alpha^{\prime}\right)^{k-1}2^{2\left(k+1\right)}k\bar{c}_{k}\bar{H}^{2k-1},\nonumber \\
\bar{g}\left(\bar{H}\right) & = & d\stackrel[k=1]{\infty}{\sum}\left(-\alpha^{\prime}\right)^{k-1}2^{2k+1}\left(2k-1\right)\bar{c}_{k}\bar{H}^{2k}.\label{eq:EOM fh gh}
\end{eqnarray}

\noindent In these EOM (\ref{eq:EOM}), if we start with the non-perturbative
and non-singular dilaton solution $\Phi$ obtained from the perturbative
solution, we can obtain $\bar{f}\left(\bar{H}\right)$ and $\bar{g}\left(\bar{H}\right)$
directly. Since the non-perturbative expansions (\ref{eq:EOM fh gh})
introduces an extra constraint $\dot{\bar{g}}=\bar{H}\dot{\bar{f}}$,
we can also determine $\bar{H}$ in the subsequent step. In other
words, all solutions can be derived from $\Phi$. However, when considering
a non-trivial Kalb-Ramond field, we can only utilize the complicated
EOM (\ref{eq:EOM1}) to (\ref{eq:EOM3}). The solution $\Phi$ in
these EOM cannot determine the solution of the matrix $\mathcal{S}$
due to equation (\ref{eq:EOM2}), which includes an infinite summation
of matrices, and we cannot extract any constraints such as $\dot{\bar{g}}=\bar{H}\dot{\bar{f}}$
from equations (\ref{eq:EOM1}) to (\ref{eq:EOM3}). The solution
$\Phi$ in these EOM cannot determine the solution of the matrix $\mathcal{S}$
due to the equation (\ref{eq:EOM2}) includes infinite summation of
matrices, and we can not extract any constraint such as $\dot{\bar{g}}=\bar{H}\dot{\bar{f}}$
from the (\ref{eq:EOM1}) to (\ref{eq:EOM3}). Therefore, it seems
impossible to obtain the non-perturbative solution including the Kalb-Ramond
field.

In this paper, our aim is to develop a method for calculating the
non-perturbative and non-singular solutions of the Hohm-Zwiebach action,
including a non-trivial Kalb-Ramond field. We begin with the three-dimensional
black string solution with axion charge \cite{Mandal:1991tz,Horne:1991gn}.
When $\left|Q\right|=M$ and we set $\lambda\rightarrow\infty$ of
cosmological constant $8/\lambda$, the metric becomes extremal, which
corresponds to the fields outside the fields outside a fundamental
macroscopic string \cite{Dabholkar:1990yf}:

\begin{equation}
ds^{2}=\frac{1}{y}\left(-dt^{2}+dx^{2}\right)+dy^{2},\label{eq:ex metric}
\end{equation}

\noindent This extremal metric includes a non-trivial Kalb-Ramond
field and exhibits a naked singularity. Our method aims to remove
this singularity, and it can be summarized as follows: At first, we
extract the ansatz based on the extremal metric (\ref{eq:ex metric})
and derive the corresponding Hohm-Zwiebach action. Secondly, we calculate
the perturbative solution up to the first order of $\alpha^{\prime}$
correction of the action. The key point in this step is that, when
we integrate out $d/dy$ from the matrix different equation (\ref{eq:EOM2}),
the constant matrix solution may also receive $\alpha^{\prime}$ corrections.
Furthermore, we observe that in our perturbative solution, the dilaton
does not receive $\alpha^{\prime}$ corrections; only the metric and
the Kalb-Ramond field are affected. Thirdly, we use a trick to transform
the infinite summation of the matrix differential equation (\ref{eq:EOM2})
into a simple trace equation:

\begin{equation}
4e^{-\Phi}\left[\stackrel[k=1]{\infty}{\sum}\left(\alpha^{\prime}\right)^{k-1}kc_{k}\mathrm{Tr}\left(\dot{\mathcal{S}}\right)^{2k}\right]=\mathrm{Tr}\left(\mathbb{C}\mathcal{S}^{-1}\dot{\mathcal{S}}\right),\label{eq:New EOM2}
\end{equation}

\noindent where $\mathbb{C}$ is $4\times4$ constant matrix. Reconsidering
the EOM (\ref{eq:EOM1}), (\ref{eq:EOM2}) and (\ref{eq:New EOM2}),
we find that all fields depend on $\Phi$, allowing us to determine
the solution of $\mathcal{S}$ through equation (\ref{eq:New EOM2}).
Finally, we obtain the regular solution of these EOM, and it implies
that the naked curvature singularity of three-dimensional extremal
black string solution can be successfully removed. As expected, the
regular solution matches the perturbative solution in the perturbative
limit $\alpha^{\prime}\rightarrow0$. While our regular solution currently
matches only the first two orders of the perturbative solution, it
can be easily generalized to arbitrary orders using our previous method
\cite{Wang:2019dcj}. 

The remainder of this paper is outlined as follows: In section 2,
we briefly review the three-dimensional extremal black string at the
tree-level. In section 3, we calculate the non-perturbative and non-singular
extremal black hole solution, including Kalb-Ramond field. Section
4 is a conclusion.

\section{Three-dimensional extremal black string at tree-level}

In this section, we demonstrate how to obtain the three-dimensional
extremal black string solution from the tree-level low energy effective
action of bosonic closed string theory. We start with the three-dimensional
low energy effective action given by:

\begin{equation}
I_{0}=\int d^{3}x\sqrt{-g}e^{-2\phi}\left(R+4\left(\partial\phi\right)^{2}-\frac{1}{12}H^{2}\right),\label{eq:effective action}
\end{equation}

\noindent where $R$ is the Ricci scalar for the metric $g_{\mu\nu}$,
$\phi$ represents the dilaton, and $H_{\mu\nu\rho}=3\partial_{\left[\mu\right.}b_{\left.\nu\rho\right]}$
is the field strength of the anti-symmetric Kalb-Ramond field $b$.
Next, we consider the following ansatz:

\begin{equation}
ds^{2}=a\left(y\right)^{2}\left(-dt^{2}+dx^{2}\right)+dy^{2}.\label{eq:ansatz}
\end{equation}

\noindent This ansatz can also be written as:

\begin{equation}
g_{\mu\nu}=\left(\begin{array}{ccc}
-a^{2}\left(y\right) & 0 & 0\\
0 & a^{2}\left(y\right) & 0\\
0 & 0 & 1
\end{array}\right),\qquad b_{\mu\nu}=\left(\begin{array}{ccc}
0 & b_{tx}\left(y\right) & 0\\
b_{xt}\left(y\right) & 0 & 0\\
0 & 0 & 0
\end{array}\right),\qquad\phi=\phi\left(y\right).
\end{equation}

\noindent It is important to note that $b_{tx}\left(y\right)=-b_{xt}\left(y\right)$.
Since the ansatz (\ref{eq:ansatz}) is independent of $y$, we can
introduce the following notation to manifest the $O\left(2,2\right)$
symmetry:

\begin{equation}
g_{\mu\nu}=\left(\begin{array}{cc}
G_{ij}\left(y\right) & 0\\
0 & 1
\end{array}\right),\qquad b_{\mu\nu}=\left(\begin{array}{cc}
B_{ij}\left(y\right) & 0\\
0 & 0
\end{array}\right),\label{eq:set up 1}
\end{equation}

\noindent where

\begin{equation}
G_{ij}\left(y\right)\equiv\left(\begin{array}{cc}
-a^{2}\left(y\right) & 0\\
0 & a^{2}\left(y\right)
\end{array}\right),\qquad B_{ij}\left(y\right)\equiv\left(\begin{array}{cc}
0 & b_{tx}\left(y\right)\\
b_{xt}\left(y\right) & 0
\end{array}\right).\label{eq:set up 2}
\end{equation}

\noindent With this notation, the action (\ref{eq:effective action})
can be rewritten as:

\begin{equation}
I_{0}=-\int dx^{3}e^{-\Phi}\left(-\dot{\Phi}^{2}-\frac{1}{8}\mathrm{Tr}\left(\dot{\mathcal{S}}^{2}\right)\right),\label{eq:tree ODD}
\end{equation}

\noindent where the dot denotes a derivative with respect to $y$,
i.e., $\dot{f}\left(y\right)\equiv\partial_{y}f\left(y\right)$, and

\begin{equation}
\mathcal{S}=\left(\begin{array}{cc}
BG^{-1} & G-BG^{-1}B\\
G^{-1} & -G^{-1}B
\end{array}\right).
\end{equation}

\noindent The $O\left(d,d\right)$ invariant dilaton $\Phi$ in action
(\ref{eq:tree ODD}) is defined in the action as:

\begin{equation}
\Phi=2\phi-\frac{1}{2}\ln\left|\det G_{ij}\right|.
\end{equation}

\noindent The EOM for the action (\ref{eq:tree ODD}) are as follows:

\begin{eqnarray}
\ddot{\Phi}+\frac{1}{8}\mathrm{Tr}\left(\dot{\mathcal{S}}\right)^{2} & = & 0,\nonumber \\
\frac{d}{dy}\left(e^{-\Phi}\mathcal{S}\left(\dot{\mathcal{S}}\right)\right) & = & 0,\nonumber \\
\dot{\Phi}^{2}+\frac{1}{8}\mathrm{Tr}\left(\dot{\mathcal{S}}\right)^{2} & = & 0.
\end{eqnarray}

\noindent The simplest solution that includes a non-trivial Kalb-Ramond
field is:

\noindent 
\begin{equation}
\Phi=0,\qquad a\left(y\right)=\frac{1}{\sqrt{y}},\qquad b_{tx}\left(y\right)=\frac{1}{y}.\label{eq:tree solution}
\end{equation}

\noindent The corresponding Kretschmann scalar is given by: 

\begin{equation}
R_{\mu\nu\rho\sigma}R^{\mu\nu\rho\sigma}=\frac{4\left(2a^{2}\ddot{a}^{2}+\dot{a}^{4}\right)}{a^{4}}=\frac{19}{4y^{4}}.
\end{equation}

\noindent It implies that the solution possesses a curvature singularity
at $y=0$ and no event horizon. This solution describes the fields
outside a fundamental macroscopic string, which corresponds to a three-dimensional
extremal black string. To understand the relationship between this
tree-level singular solution (\ref{eq:tree solution}) and the extremal
black string, let us recall the three-dimensional black string with
an axion charge:

\begin{equation}
ds^{2}=-\left(1-\frac{M}{r}\right)dt^{2}+\left(1-\frac{Q^{2}}{Mr}\right)dx^{2}+\left(1-\frac{M}{r}\right)^{-1}\left(1-\frac{Q^{2}}{Mr}\right)^{-1}\frac{\lambda dr^{2}}{8r^{2}},
\end{equation}

\noindent with

\begin{equation}
H_{rtx}=\frac{Q}{r^{2}},\qquad\phi=\ln r+\frac{1}{2}\ln\frac{\lambda}{2},
\end{equation}

\noindent where $8/\lambda$ is related to the cosmological constant.
The extremal metric is obtained by setting $\left|Q\right|=M$, which
gives:

\begin{equation}
ds^{2}=\left(1-\frac{M}{r}\right)\left(-dt^{2}+dx^{2}\right)+\left(1-\frac{M}{r}\right)^{-2}\frac{\lambda dr^{2}}{8r^{2}}.\label{eq:extremal metric}
\end{equation}

\noindent Using the coordinate transformations:

\begin{equation}
\tilde{y}=-\left(\frac{\lambda}{8}\right)^{1/2}\frac{r}{M},\qquad\tilde{t}=\left(\frac{\lambda}{8}\right)^{1/4}x,\qquad\tilde{x}=\left(\frac{\lambda}{8}\right)^{1/4}t,
\end{equation}

\noindent and taking the limit $\lambda\rightarrow\infty$, which
corresponds to the vanishing cosmological constant in the action (\ref{eq:effective action}),
the extremal metric (\ref{eq:extremal metric}) becomes:

\begin{equation}
ds^{2}=\frac{1}{\tilde{y}}\left(-d\tilde{t}^{2}+d\tilde{x}^{2}\right)+d\tilde{y}^{2},
\end{equation}

\noindent which matches the solution (\ref{eq:tree solution}). In
this ansatz, we can employ the isotropic Hohm-Zwiebach action without
considering the multitrace terms.

\section{Regular extremal black string via $\alpha^{\prime}$ corrections}

In this section, our aim is to remove the naked singularity of the
solution (\ref{eq:tree solution}) by using the Hohm-Zwiebach action.
Hohm and Zwiebach demonstrated that the following low energy effective
action with complete $\alpha^{\prime}$ corrections can be expressed
as

\begin{eqnarray}
I_{HZ} & = & \int d^{3}x\sqrt{-g}e^{-2\phi}\left(R+4\left(\partial\phi\right)^{2}-\frac{1}{12}H^{2}+\frac{1}{4}\alpha^{\prime}\left(R^{\mu\nu\rho\sigma}R_{\mu\nu\rho\sigma}+\ldots\right)+\alpha^{\prime2}\left(\ldots\right)+\ldots\right)\nonumber \\
 & = & -\int dx^{3}e^{-\Phi}\left(-\dot{\Phi}^{2}+\stackrel[k=1]{\infty}{\sum}\left(\alpha^{\prime}\right)^{k-1}\bar{c}_{k}\mathrm{Tr}\left(\dot{\mathcal{S}}^{2k}\right)\right),\label{eq:corrected action}
\end{eqnarray}

\noindent where $\bar{c}_{1}=c_{1}=-\frac{1}{8}$, $\bar{c}_{2}=-c_{2}=-\frac{1}{64}$,
$\bar{c}_{3}=c_{3}=-\frac{1}{3.2^{7}}$, $\bar{c}_{4}=-c_{4}=-\frac{1}{2^{15}}+\frac{1}{2^{12}}\zeta\left(3\right)$
and $\bar{c}_{k>4}$'s are unknown coefficients for the bosonic case
\cite{Codina:2021cxh}. The EOM of the Hohm-Zwiebach action are given
by:

\begin{eqnarray}
\ddot{\Phi}-\stackrel[k=1]{\infty}{\sum}\left(\alpha^{\prime}\right)^{k-1}k\bar{c}_{k}\mathrm{Tr}\left(\dot{\mathcal{S}}\right)^{2k} & = & 0,\nonumber \\
\frac{d}{dy}\left(e^{-\Phi}\stackrel[k=1]{\infty}{\sum}\left(\alpha^{\prime}\right)^{k-1}4k\bar{c}_{k}\mathcal{S}\left(\dot{\mathcal{S}}\right)^{2k-1}\right) & = & 0,\nonumber \\
\dot{\Phi}^{2}-\stackrel[k=1]{\infty}{\sum}\left(\alpha^{\prime}\right)^{k-1}\left(2k-1\right)\bar{c}_{k}\mathrm{Tr}\left(\dot{\mathcal{S}}\right)^{2k} & = & 0.\label{eq:HZ EOM}
\end{eqnarray}

\noindent To calculate the perturbative solution, let us introduce
a new variable $\Omega\equiv e^{-\Phi}$, where $\dot{\Omega}=-\dot{\Phi}\Omega$
and $\ddot{\Omega}=\left(-\ddot{\Phi}+\dot{\Phi}^{2}\right)\Omega$.
The EOM can be rewritten as:

\begin{eqnarray}
\frac{\left(\dot{\Omega}\right)^{2}}{\Omega^{2}}-\frac{\ddot{\Omega}}{\Omega}-\stackrel[k=1]{\infty}{\sum}\left(\alpha^{\prime}\right)^{k-1}kc_{k}\mathrm{Tr}\left(\dot{\mathcal{S}}\right)^{2k} & = & 0,\nonumber \\
\frac{d}{dy}\left(e^{-\Phi}\stackrel[k=1]{\infty}{\sum}\left(\alpha^{\prime}\right)^{k-1}4kc_{k}\mathcal{S}\left(\dot{\mathcal{S}}\right)^{2k-1}\right) & = & 0,\nonumber \\
\frac{\dot{\Omega}^{2}}{\Omega^{2}}-\stackrel[k=1]{\infty}{\sum}\left(\alpha^{\prime}\right)^{k-1}\left(2k-1\right)c_{k}\mathrm{Tr}\left(\dot{\mathcal{S}}\right)^{2k} & = & 0,\label{eq:per EOM}
\end{eqnarray}

\noindent Next, we make the assumption that the perturbative solutions
of the EOM given by (\ref{eq:pertur form}) take the following forms:

\begin{eqnarray}
\Omega\left(y\right) & = & \Omega_{0}\left(y\right)+\alpha^{\prime}\Omega_{1}\left(y\right)+\alpha^{\prime2}\Omega_{2}\left(y\right)+\ldots,\nonumber \\
a\left(r\right) & = & a_{0}\left(r\right)+\alpha^{\prime}a_{1}\left(r\right)+\alpha^{\prime2}a_{2}\left(r\right)+\ldots.\nonumber \\
b\left(r\right) & = & b_{0}\left(r\right)+\alpha^{\prime}b_{1}\left(r\right)+\alpha^{\prime2}b_{2}\left(r\right)+\ldots,\label{eq:pertur form}
\end{eqnarray}

\noindent where we define $b_{tx}\left(y\right)\equiv b\left(y\right)$.
Substituting the perturbative forms (\ref{eq:pertur form}) into the
EOM (\ref{eq:pertur form}) and considering the zeroth order in $\alpha^{\prime}$,
we obtain the following equations:

\begin{eqnarray}
\ddot{\Omega}_{0}\left(y\right) & = & 0,\nonumber \\
\dot{\Omega}_{0}^{2}+\frac{\Omega_{0}^{2}\left(\dot{b}_{0}^{2}-4a_{0}^{2}\dot{a}_{0}^{2}\right)}{2a_{0}^{4}} & = & 0,
\end{eqnarray}

\noindent and the second equation of the EOM (\ref{eq:pertur form})
leads to the following expression:

\begin{equation}
\frac{d}{dy}\left(\begin{array}{cccc}
\mathbb{M}_{11}^{0} & 0 & 0 & \mathbb{M}_{14}^{0}\\
0 & \mathbb{M}_{11}^{0} & -\mathbb{M}_{14}^{0} & 0\\
0 & -\mathbb{M}_{41}^{0} & \mathbb{M}_{44}^{0} & 0\\
\mathbb{M}_{41}^{0} & 0 & 0 & \mathbb{M}_{44}^{0}
\end{array}\right)=0,
\end{equation}

\noindent where the entries of the matrix are given by

\begin{eqnarray}
\mathbb{M}_{11}^{0} & = & \frac{\Omega_{0}\left(2a_{0}^{3}\dot{a}_{0}-b_{0}\dot{b}_{0}\right)}{2a_{0}^{4}},\nonumber \\
\mathbb{M}_{41}^{0} & = & \frac{\Omega_{0}\dot{b}_{0}}{2a_{0}^{4}},\nonumber \\
\mathbb{M}_{14}^{0} & = & \frac{\Omega_{0}\left(4a_{0}^{3}\dot{a}_{0}b-\left(a_{0}^{4}+b_{0}^{2}\right)\dot{b}_{0}\right)}{2a_{0}^{4}},\nonumber \\
\mathbb{M}_{44}^{0} & = & \frac{\Omega_{0}\left(b_{0}\dot{b}_{0}-2a_{0}^{3}\dot{a}_{0}\right)}{2a_{0}^{4}}.
\end{eqnarray}

\noindent The solution is

\begin{equation}
\Omega_{0}=1,\qquad a_{0}=\frac{1}{\sqrt{y}},\qquad b_{0}=\frac{1}{y},
\end{equation}

\noindent which is consistent with the solution (\ref{eq:tree solution}).
The integration of $d/dy$ in the second equation of EOM (\ref{eq:per EOM})
yields the constant matrix

\begin{equation}
\left(\begin{array}{cccc}
\mathbb{M}_{11}^{0} & 0 & 0 & \mathbb{M}_{14}^{0}\\
0 & \mathbb{M}_{11}^{0} & -\mathbb{M}_{14}^{0} & 0\\
0 & -\mathbb{M}_{41}^{0} & \mathbb{M}_{44}^{0} & 0\\
\mathbb{M}_{41}^{0} & 0 & 0 & \mathbb{M}_{44}^{0}
\end{array}\right)=\left(\begin{array}{cccc}
0 & 0 & 0 & 0\\
0 & 0 & 0 & 0\\
0 & -1 & 0 & 0\\
1 & 0 & 0 & 0
\end{array}\right).\label{eq:matrix 0}
\end{equation}

\noindent This constant matrix plays a crucial role in determining
the non-perturbative solution. Considering the EOM (\ref{eq:pertur form})
at the first order of $\alpha^{\prime}$, we have:

\begin{eqnarray*}
\ddot{\Omega}_{1}+\frac{\Omega_{0}\left(\dot{b}_{0}^{2}-4a_{0}^{2}\dot{a}_{0}^{2}\right){}^{2}}{16a_{0}^{8}} & = & 0,\\
2\dot{\Omega}_{0}\dot{\Omega}_{1}+\frac{\Omega_{0}\Omega_{1}\left(\dot{b}_{0}^{2}-4a_{0}^{2}\dot{a}_{0}^{2}\right)}{a_{0}^{4}}\\
+\Omega_{0}^{2}\left(\frac{3\left(\dot{b}_{0}^{2}-4a_{0}^{2}\dot{a}_{0}^{2}\right){}^{2}}{16a_{0}^{8}}+\frac{a_{0}\dot{b}_{0}\dot{b}_{1}-2a_{1}\dot{b}_{0}^{2}-4a_{0}^{3}\dot{a}_{0}\dot{a}_{1}+4a_{1}a_{0}^{2}\dot{a}_{0}^{2}}{a_{0}^{5}}\right) & = & 0,
\end{eqnarray*}

\noindent and

\begin{equation}
\frac{d}{dy}\left(\begin{array}{cccc}
\mathbb{M}_{11}^{1} & 0 & 0 & \mathbb{M}_{14}^{1}\\
0 & \mathbb{M}_{11}^{1} & -\mathbb{M}_{14}^{1} & 0\\
0 & -\mathbb{M}_{41}^{1} & \mathbb{M}_{44}^{1} & 0\\
\mathbb{M}_{41}^{1} & 0 & 0 & \mathbb{M}_{44}^{1}
\end{array}\right)=0,
\end{equation}

\noindent where

\begin{eqnarray}
\mathbb{M}_{11}^{1} & = & -\frac{1}{8a_{0}^{8}}\left[4a_{0}^{4}\left(\left(b_{1}\Omega_{0}+b_{0}\Omega_{1}\right)\dot{b}_{0}+\Omega_{0}b_{0}\dot{b}_{1}\right)-2a_{0}^{3}\Omega_{0}\dot{b}_{0}\left(\dot{a}_{0}\dot{b}_{0}+8a_{1}b_{0}\right)-4\Omega_{0}a_{0}^{2}\dot{a}_{0}^{2}b_{0}\dot{b}_{0}\right.\nonumber \\
 &  & \left.-8a_{0}^{7}\left(\Omega_{1}\dot{a}_{0}+\Omega_{0}\dot{a}_{1}\right)+8\Omega_{0}a_{0}^{6}\dot{a}_{0}a_{1}+8\Omega_{0}a_{0}^{5}\dot{a}_{0}^{3}+\Omega_{0}b_{0}\dot{b}_{0}^{3}\right],\nonumber \\
\mathbb{M}_{41}^{1} & = & \frac{1}{8a_{0}^{8}}\left[4\Omega_{0}a_{0}^{4}\dot{b}_{1}+4a_{0}^{2}\dot{b}_{0}\left(-\Omega_{0}\dot{a}_{0}^{2}+a_{0}^{2}\Omega_{1}-4\Omega_{0}a_{1}a_{0}\right)+\Omega_{0}(y)\dot{b}_{0}^{3}\right],\nonumber \\
\mathbb{M}_{14}^{1} & = & -\frac{1}{8a_{0}^{8}}\left[4a_{0}^{8}\left(\Omega_{1}\dot{b}_{0}+\Omega_{0}\dot{b}_{1}\right)-16a_{0}^{7}\left(\dot{a}_{0}\left(\Omega_{0}b_{1}+\Omega_{1}b_{0}\right)+\Omega_{0}\dot{a}_{1}b_{0}\right)-4a_{0}^{6}\Omega_{0}\dot{a}_{0}\left(\dot{a}_{0}\dot{b}_{0}-4a_{1}b_{0}\right)\right.\nonumber \\
 &  & +16\Omega_{0}a_{0}^{5}\dot{a}_{0}^{3}b_{0}+a_{0}^{4}\left(\Omega_{0}\dot{b}_{0}^{3}+4b_{0}\left(2\Omega_{0}b_{1}+\Omega_{1}b_{0}\right)\dot{b}_{0}+4\Omega_{0}b_{1}^{2}\dot{b}_{1}\right)\nonumber \\
 &  & \left.-4\Omega_{0}a_{0}^{3}b_{0}\dot{b}_{0}\left(\dot{a}_{0}\dot{b}_{0}+4a_{1}b_{0}\right)-4\Omega_{0}a_{0}^{2}\dot{a}_{0}^{2}b_{0}^{2}\dot{b}_{0}+\Omega_{0}b_{0}^{2}\dot{b}_{0}^{3}\right],\nonumber \\
\mathbb{M}_{44}^{1} & = & \frac{1}{8a_{0}^{8}}\left[4a_{0}^{4}\left(\left(\Omega_{0}b_{1}+\Omega_{1}b_{0}\right)\dot{b}_{0}+\Omega_{0}b_{0}\dot{b}_{1}\right)-2\Omega_{0}a_{0}^{3}\dot{b}_{0}\left(\dot{a}_{0}\dot{b}_{0}+8a_{1}b_{0}\right)-4\Omega_{0}a_{0}^{2}\dot{a}_{0}^{2}b_{0}\dot{b}_{0}\right.\nonumber \\
 &  & \left.-8a_{0}^{7}\left(\Omega_{1}\dot{a}_{0}+\Omega_{0}\dot{a}_{1}\right)+8\Omega_{0}a_{0}^{6}\dot{a}_{0}a_{1}+8\Omega_{0}a_{0}^{5}\dot{a}_{0}^{3}+\Omega_{0}b_{0}\dot{b}_{0}^{3}\right].
\end{eqnarray}

\noindent The solution is

\begin{equation}
\Omega_{1}=0,\qquad a_{0}=-\frac{C}{y^{3/2}},\qquad b_{0}=-\frac{2C}{y^{2}}.
\end{equation}

\noindent where $C$ is a non-zero constant and

\begin{equation}
\left(\begin{array}{cccc}
\mathbb{M}_{11}^{1} & 0 & 0 & \mathbb{M}_{14}^{1}\\
0 & \mathbb{M}_{11}^{1} & -\mathbb{M}_{14}^{1} & 0\\
0 & -\mathbb{M}_{41}^{1} & \mathbb{M}_{44}^{1} & 0\\
\mathbb{M}_{41}^{1} & 0 & 0 & \mathbb{M}_{44}^{1}
\end{array}\right)=\left(\begin{array}{cccc}
0 & 0 & 0 & 0\\
0 & 0 & 0 & 0\\
0 & 0 & 0 & 0\\
0 & 0 & 0 & 0
\end{array}\right).\label{eq:matrix 1}
\end{equation}

\noindent Therefore, the perturbative solution, including the first
two orders of $\alpha^{\prime}$, can be expressed as:

\begin{eqnarray}
a\left(y\right) & = & \frac{1}{\sqrt{y}}-\frac{C}{y^{3/2}}\alpha^{\prime}+\cdots,\nonumber \\
b\left(y\right) & = & \frac{1}{y}-\frac{2C}{y^{2}}\alpha^{\prime}+\cdots,\nonumber \\
\Omega\left(y\right) & = & 1+0\cdot\alpha^{\prime}+\cdots.\label{eq:perturbed solution}
\end{eqnarray}

Next, let us proceed with the calculation of the non-singular and
non-perturbative solution. To do it, we can rewrite the EOM (\ref{eq:per EOM})
as follows:

\begin{eqnarray}
\ddot{\Phi}-y\left(\mathcal{S}\right) & = & 0,\nonumber \\
\frac{d}{dy}\left(e^{-\Phi}\stackrel[k=1]{\infty}{\sum}\left(\alpha^{\prime}\right)^{k-1}4kc_{k}\mathcal{S}\left(\dot{\mathcal{S}}\right)^{2k-1}\right) & = & 0,\nonumber \\
\dot{\Phi}^{2}+g\left(\mathcal{S}\right) & = & 0,\label{eq:non-per EOM}
\end{eqnarray}

\noindent where

\begin{eqnarray}
y\left(\mathcal{S}\right) & = & \stackrel[k=1]{\infty}{\sum}\left(\alpha^{\prime}\right)^{k-1}kc_{k}\mathrm{Tr}\left(\dot{\mathcal{S}}\right)^{2k},\nonumber \\
g\left(\mathcal{S}\right) & = & -\stackrel[k=1]{\infty}{\sum}\left(\alpha^{\prime}\right)^{k-1}\left(2k-1\right)c_{k}\mathrm{Tr}\left(\dot{\mathcal{S}}\right)^{2k}.
\end{eqnarray}

\noindent Based on the perturbative solution (\ref{eq:perturbed solution}),
we can determine the non-perturbative solution for the $O\left(d,d\right)$
dilaton:

\begin{equation}
\Phi=0,
\end{equation}

\noindent which covers the perturbative solution (\ref{eq:perturbed solution})
as $\alpha^{\prime}\rightarrow0$. From the first and second equations
of (\ref{eq:non-per EOM}), we obtain:

\begin{equation}
y\left(\mathcal{S}\right)=g\left(\mathcal{S}\right)=0.
\end{equation}

\noindent Hence, the crucial equation is the second one in (\ref{eq:non-per EOM}).
However, this equation represents a differential equation involving
an infinite summation of matrices. Finding the solution for $\mathcal{S}$
seems impossible in this case. To make progress, we use a trick here.
Let us begin with the second equation in (\ref{eq:non-per EOM})

\begin{equation}
\frac{d}{dy}\left(e^{-\Phi}\stackrel[k=1]{\infty}{\sum}\left(\alpha^{\prime}\right)^{k-1}4kc_{k}\mathcal{S}\left(\dot{\mathcal{S}}\right)^{2k-1}\right)=0.
\end{equation}

\noindent After integrating out $d/dy$, we obtain:

\begin{equation}
e^{-\Phi}\stackrel[k=1]{\infty}{\sum}\left(\alpha^{\prime}\right)^{k-1}4kc_{k}\mathcal{S}\left(\dot{\mathcal{S}}\right)^{2k-1}=\mathbb{C},
\end{equation}

\noindent where $\mathbb{C}$ is a $4\times4$ constant matrix. Multiplying
both sides by $\mathcal{S}^{-1}$ and then $\dot{\mathcal{S}}$, we
have:

\begin{equation}
e^{-\Phi}\stackrel[k=1]{\infty}{\sum}\left(\alpha^{\prime}\right)^{k-1}4kc_{k}\left(\dot{\mathcal{S}}\right)^{2k}=\mathbb{C}\mathcal{S}^{-1}\dot{\mathcal{S}}.
\end{equation}

\noindent Finally, by taking the trace on both sides of the equation,
we obtain:

\begin{equation}
4e^{-\Phi}y\left(\mathcal{S}\right)=\mathrm{Tr}\left(\mathbb{C}\mathcal{S}^{-1}\dot{\mathcal{S}}\right).\label{eq:trick equation}
\end{equation}

\noindent Therefore, the infinite summation of matrices is reduced
to a finite-term differential equation. To solve this equation, we
follow two steps. The first step involves guessing the regular solution
for the $b$ field, which is

\begin{equation}
b\left(y\right)=\frac{1}{y+2\alpha^{\prime}C}.
\end{equation}

\noindent It covers the perturbative solution (\ref{eq:perturbed solution})
as $\alpha^{\prime}\rightarrow0$. The next step is to determine the
constant matrix $\mathbb{C}$. Based on the previous results (\ref{eq:matrix 0})
and (\ref{eq:matrix 1}), we can assign the constant matrix $\mathbb{C}$
as

\begin{equation}
\mathbb{C}=\left(\begin{array}{cccc}
0 & 0 & 0 & 0\\
0 & 0 & 0 & 0\\
0 & -1 & 0 & 0\\
1 & 0 & 0 & 0
\end{array}\right).
\end{equation}

\noindent Consequently, the equation (\ref{eq:trick equation}) simplifies
to

\begin{equation}
2+2a^{3}\left(y+2\alpha^{\prime}C\right)^{2}\left(a+4\dot{a}\left(y+2\alpha^{\prime}C\right)\right)=0.
\end{equation}

\noindent The solution of this equation is

\begin{equation}
a\left(y\right)=\frac{1}{\sqrt{y+2\alpha^{\prime}C}}.
\end{equation}

\noindent In summary, we have successfully derived non-perturbative
solutions that satisfy the EOM (\ref{eq:HZ EOM}):

\begin{equation}
ds^{2}=a\left(y\right)^{2}\left(-dt^{2}+dx^{2}\right)+dy^{2},
\end{equation}

\noindent with

\begin{equation}
\Phi=0,\qquad a\left(y\right)=\frac{1}{\sqrt{y+2\alpha^{\prime}C}},\qquad b\left(y\right)=\frac{1}{y+2\alpha^{\prime}C},
\end{equation}

\begin{figure}[H]
\begin{centering}
\includegraphics[scale=0.5]{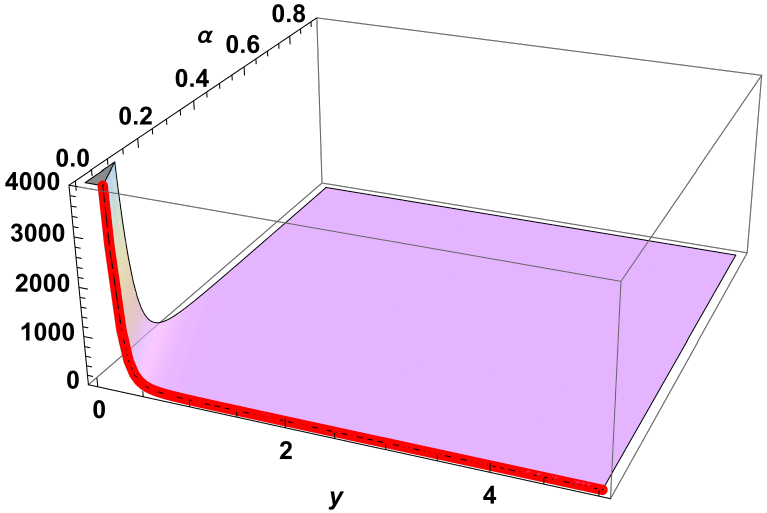} $\qquad\qquad$\includegraphics[scale=0.5]{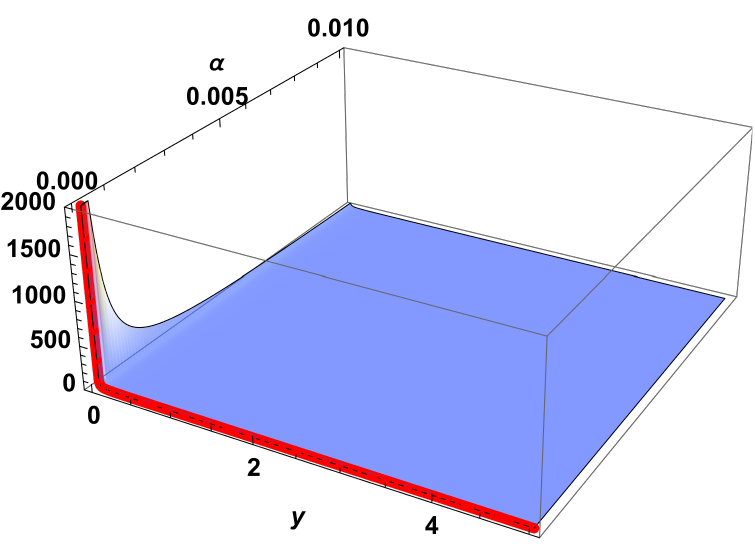}
\par\end{centering}
\begin{centering}
$\qquad\qquad$$\qquad\qquad$$\qquad\qquad$$\qquad\qquad$ 
\par\end{centering}
\begin{centering}
$R_{\mu\nu\rho\sigma}R^{\mu\nu\rho\sigma}\left(y,\alpha^{\prime}\right)$$\qquad\qquad$$\qquad\qquad$$\qquad\qquad$$\qquad\qquad$
$b\left(y,\alpha^{\prime}\right)$
\par\end{centering}
\centering{}\caption{\label{fig:RB} The figures display $R_{\mu\nu\rho\sigma}R^{\mu\nu\rho\sigma}\left(y,\alpha^{\prime}\right)$
and $b\left(y,\alpha^{\prime}\right)$, with the assumption $C=1$.
The red lines represent the singular solutions $R_{\mu\nu\rho\sigma}R^{\mu\nu\rho\sigma}\left(y,0\right)$
and $b\left(y,0\right)$ obtained from (\ref{eq:tree solution}).
The singularities are located at $y=0$ and $\alpha^{\prime}=0$.}
\end{figure}

\noindent It is straightforward to verify that this solution matches
with the perturbative solution as $\alpha^{\prime}\rightarrow0$.
Although our regular solution matches only the first two orders of
the perturbative solution, it can be easily generalized to arbitrary
orders using our previous method \cite{Wang:2019dcj}. Using this
solution, we can calculate the Kretschmann scalar: 

\begin{equation}
R_{\mu\nu\rho\sigma}R^{\mu\nu\rho\sigma}=\frac{4\left(2a^{2}\ddot{a}^{2}+\dot{a}^{4}\right)}{a^{4}}=\frac{19}{4\left(y+2\alpha^{\prime}C\right){}^{4}}.
\end{equation}

\noindent which is regular for arbitrary $y\geq0$. To illustrate
how $\alpha^{\prime}$ corrections affects the naked singularity,
we provide a representation in Figure (\ref{fig:RB}). It is evident
that once the $\alpha^{\prime}$ corrections are introduced, the singularities
in the metric and Kalb-Ramond field disappear.

\section{Conclusion }

In this paper, we have successfully calculated the non-perturbative
and non-singular solution for the Hohm-Zwiebach action, considered
the presence of a non-trivial Kalb-Ramond field. In order to calculate
the matrix differential equations, we transformed the infinite summation
of matrices into a simple trace of a matrix, enabled us to solve the
matrix differential equations. As a result, we obtained regular solutions
for all the fields, including the spacetime metric. This implies that
the naked singularity of the extremal black string can be eliminated
by the $\alpha^{\prime}$ corrections. In our future work, we plan
to study the following aspects:
\begin{itemize}
\item Extension to anisotropic backgrounds: In this study, we focused on
the isotropic background. However, it would be interesting to investigate
the anisotropic Hohm-Zwiebach action. It implies that we need to consider
the multitrace terms in the Hohm-Zwiebach action and refine our method
accordingly.
\item Application to BTZ black hole: The BTZ black hole solution in string
theory also requires a non-trivial Kalb-Ramond field. Therefore, if
we aim to address the curvature singularity in the BTZ black hole,
it is crucial to incorporate the Kalb-Ramond field in the Hohm-Zwiebach
action.
\end{itemize}
By addressing these directions, we can further broaden our understanding
of non-perturbative solutions and their implications for resolving
singularities in black hole spacetimes.

\bigskip 

\vspace{5mm}

\noindent {\bf Acknowledgements} 
We are very grateful to Xin Li, Peng Wang, Houwen Wu and Haitang Yang for many illuminating discussions and suggestions. This work is supported in part by NSFC (Grant No. 12105031), and the Postdoctoral Science Foundation of Chongqing (Grant No. cstc2021jcyj-bshX0227).

\end{document}